\documentclass[preprint,12pt,authoryear]{elsarticle}

\usepackage{amssymb}
\usepackage{amsmath}
\journal{Planetary and Space Science}

\begin{document}

\begin{frontmatter}

%% Title, authors and addresses

%% use the tnoteref command within \title for footnotes;
%% use the tnotetext command for theassociated footnote;
%% use the fnref command within \author or \affiliation for footnotes;
%% use the fntext command for theassociated footnote;
%% use the corref command within \author for corresponding author footnotes;
%% use the cortext command for theassociated footnote;
%% use the ead command for the email address,
%% and the form \ead[url] for the home page:
%% \title{Title\tnoteref{label1}}
%% \tnotetext[label1]{}
%% \author{Name\corref{cor1}\fnref{label2}}
%% \ead{email address}
%% \ead[url]{home page}
%% \fntext[label2]{}
%% \cortext[cor1]{}
%% \affiliation{organization={},
%%            addressline={}, 
%%            city={},
%%            postcode={}, 
%%            state={},
%%            country={}}
%% \fntext[label3]{}

\title{A discussion on estimating small bodies taxonomies using phase curves results} %% Article title

%% use optional labels to link authors explicitly to addresses:
%% \author[label1,label2]{}
%% \affiliation[label1]{organization={},
%%             addressline={},
%%             city={},
%%             postcode={},
%%             state={},
%%             country={}}
%%
%% \affiliation[label2]{organization={},
%%             addressline={},
%%             city={},
%%             postcode={},
%%             state={},
%%             country={}}

\author{Alvaro Alvarez-Candal} %% Author name

%% Author affiliation
\affiliation{organization={Instituto de Astrofísica de Andalucía, CSIC},%Department and Organization
            addressline={Apt 3004}, 
            city={Granada},
            postcode={E18080}, 
            %state={},
            country={Spain}}
\affiliation{organization={Instituto de Física Aplicada a las Ciencias y las Tecnologías, Universidad de Alicante},%Department and Organization
            addressline={San Vicent del Raspeig, E03080}, 
            city={Alicante},
            postcode={E03080}, 
            %state={},
            country={Spain}}

%% Abstract
\begin{abstract}
%% Text of abstract
Upcoming large multiwavelength photometric surveys will provide a leap in our understanding of small body populations, among other fields of modern astrophysics. Serendipitous observations of small bodies in different orbital locations allow us to study diverse phenomena related to how their surfaces scatter solar light.

In particular, multiple observations of the same object in different epochs permit us to construct their phase curves to obtain absolute magnitudes and phase coefficients. In this work, we tackle a series of long-used relationships associating these phase coefficients with the taxa of small bodies and suggest that some may need to be revised in the light of large-number statistics. 
\end{abstract}

%%Graphical abstract
%\begin{graphicalabstract}
%\includegraphics{grabs}
%end{graphicalabstract}

%%Research highlights
\begin{highlights}
\item The geometric albedo vs. slope of the linear part of the phase curve relation may not hold to large numbers.
\item {Taxonomy analyses from phase coefficients should be considered with care.}
\end{highlights}

%% Keywords
\begin{keyword}
%% keywords here, in the form: keyword \sep keyword
Small Solar System bodies \sep Astronomy data analysis \sep Catalogs 

%% PACS codes here, in the form: \PACS code \sep code

%% MSC codes here, in the form: \MSC code \sep code
%% or \MSC[2008] code \sep code (2000 is the default)
\end{keyword}

\end{frontmatter}

%% Add \usepackage{lineno} before \begin{document} and uncomment 
%% following line to enable line numbers
%% \linenumbers

%% main text
%%

%% Use \section commands to start a section
\section{Introduction} \label{sec:intro}

\cite{bobrov1929} showed that asteroids do not only reflect light from the Sun but that light carries information from the surfaces, showing that not all asteroids' surfaces are equal. Nevertheless, it was only in \cite {cmz1975Icar} that two broad groups with different reflectivity, i.e., colors, the S and C, were recognized. Further photometric analyses using multi-wavelength data re-identified the C and S major groupings while adding more details and generating the first taxonomies \citep[see][]{Tholen1984PhDT,tholenbarucci1989}. The arrival of extensive photometric surveys, in particular the Sloan Digital Sky Survey's Moving Objects Catalog \citep{ivezic2001AJ,juric2002AJ}, further increased the details in the taxonomic distributions, but it kept the discrimination in two large complexes \citep[see, for example,][and references therein]{carvano2010,hasselmann2015,colazo2022}
Certainly, spectroscopic data produce even more detailed taxonomic classifications \citep[for example][]{busbinzel2002tax,demeo2009tax, mahlke2022tax} but for the scope of this work, that level of detail is too large; therefore, we will only discuss photometric data in the remaining of this work.

The taxonomic classification is a proxy for the mineralogical properties of the surface of small bodies, hence its importance. The distribution of the taxa in the Solar System helps us understand the evolution that {the Solar System suffered}, not only from a dynamical point of view but also its physical-chemical evolution \citep[for instance][]{gradietedesco1982Sci,bell1989asterois2,monthediniz2003Icar,demeo2013Icar,demeo2014Natur}. The procedure to assign taxa uses colors obtained in a single observation (usually close in time if not simultaneous) or averages (weighted by uncertainties mostly) and different clustering techniques with convenient thresholds that define the borders between different groups; nevertheless, using a single observation or an average of the colors, although a valid approach when few data are available, turns out to increase the blur in borders between taxa due to {phase coloring} \citep{alcan2024}. Thus, we have been proposing a different approach: to determine the phase curves of the objects using multi-wavelength data from large surveys \citep{alcan2022} and measure absolute colors, defined as the difference of absolute magnitudes in different filters \citep{ayala2018}, to determine taxonomical classification.

The phase curve of asteroids shows the change of apparent brightness, normalized to unit distance from the Sun and the observer, with the phase angle $(\alpha$). The angle $\alpha$ is the angular distance subtended between the Sun and the observer, as seen from the object. The phase curves usually show an increasing behavior characterized by a linear change for $\alpha\geq5-9$ deg, and a sudden increase in brightness at low-$\alpha$ (the opposition effect, \citealt{hapke1963JGR,muino1989OE}).

To make the most of the information extracted from phase curves, we need to use a photometric model that describes the behavior of the magnitude (or flux) with $\alpha$ in the generic form
\begin{equation}\label{eq1}
    M(\alpha) = H + f(\vec{G},\vec{\Phi},\alpha),
\end{equation}
where $H$ is the absolute magnitude, $\vec{G}$ are the phase coefficients, and $\vec{\Phi}$ are functions that describe the shape of the phase curve. In the simplest form, $f(\vec{G},\vec{\Phi},\alpha) = \beta\alpha$ and Eq. \ref{eq1} is reduced to a linear approximation, which is often used with transneptunian objects because of their restricted phase angle coverage ($\alpha<2$ deg). In this work we will restrict to the HG$_1$G$_2$ \citep{muinonen2010HG1G2} and the HG$^*_{12}$ \citep{penti2016HG} photometric models. The former is the IAU-accepted model for small bodies, while the latter is an adaptation to work with lower-quality data or sparse coverage of the phase curves. {A brief description of the models is given in Sect. \ref{sec:problems}.} The procedure is to fit the best values of $\vec{G}$ and $H$ from the observed apparent magnitudes and phase angles. 

The phase coefficients, which can be and have been used to estimate the taxa of asteroids, are usually linked to previous taxonomical classifications. This linkage makes sense as it provides a second use to data taken, generally, with other purposes, such as multi-epoch lightcurves or straight phase curve measurements. However, these studies are limited because the data used were taken in one or two photometric filters in traditional observing runs that maximize the observations for a limited sample of objects in a restricted span of phase angles. Yet, with the data produced by massive photometric surveys, it may be possible to use the phase curves to compute the absolute magnitudes necessary to estimate taxa without using the phase coefficients.

In the next Section, we will describe some of the limitations encountered in the literature with taxa estimation from phase curves. In contrast, in the final Section, we argue that multi-wavelength is the best approach. {Although some of the results presented here were hinted at in previous works, this is the first time these issues are put forward in one place.}

\section{Issues of estimating taxa from phase curves} \label{sec:problems}
We need some background about the photometric models before describing the issues below. As mentioned above, we will use the HG$_1$G$_2$ and HG$^*_{12}$ models. \cite{muinonen2010HG1G2} defined the first model as
\begin{equation}\label{eq2}
f(\vec{G},\vec{\Phi},\alpha)=-2.5\log{[G_1\Phi_1(\alpha)+G_2\Phi_2(\alpha)+(1-G_1-G_2)\Phi_3(\alpha)]},
\end{equation}
where the $\Phi_i$ functions are tabulated. In their work, they proposed a version using a single phase coefficient G$_{12}$ such that G$_i=$G$_i$(G$_{12}$), $i=1,2$, that worked better for a small number of observations. Later, \cite{penti2016HG} proposed an improvement over the model with a new definition of the parametric approximations  G$_i=$G$_i$(G$^*_{12}$), $i=1,2$, keeping the same form of the photometric model as shown in Eq. \ref{eq2}. Therefore, the reader should remember that G$_{12}\neq$ G$^*_{12}$.

\subsection{Data}
{Most data presented in this work was obtained following the procedures described in \cite{alcan2022} and \cite{alcan2024}, unless explicitly mentioned.

To briefly describe the processing of the data: We used as input the data from the revision of the Sloan Digital Sky Survey by \cite{sergeyev2021AyA}. From the initial database, of more than $10^6$ observations of almost $4\times10^5$ small bodies, we selected only objects with at least three observations and a minimum span of at least 5 deg in phase angle. The only quality cut applied was ignoring any data with photometric error more significant than 1 mag.

The processing of the data includes the probability distribution of the possible rotational light-curve amplitude of the object, as well as its rotational phase (see Eq. 3 in \citealt{alcan2024}), which is then used to fit the photometric model of choice, see the following sections for details. The estimation of the phase coefficients is made by the computation of the complete probability distributions unless explicitly stated. 
}

\subsection{Single parametric model: G$_{12}$ and G$_{12}^{*}$}

Since the 80s' it has been suggested that the average values of phase coefficients for different taxonomical types are different \citep{harrisyoung1989Icar} and can be used to estimate taxa. {Note that, at the time, the photometric model used was the HG model \citep{bowell1989aste}. Nevertheless,} the usual uncertainties in the average values and the spread of each taxon in the phase coefficients phase space argue against a clear relation {when analyzing massive numbers of phase curves \citep[e.g.,][]{Oszki2012Icar,mahlke2021}. Nevertheless, it must be noted that when using high-quality and well-covered phase curves of individual objects, differences in the shape of the phase curves are apparent \citep{shev2016}.}

If the phase coefficients are good estimators of the small bodies taxa, they should show some relation with either wavelength and (or) the actual taxa of the objects. Therefore, we first analyze the possible dependence of the phase coefficients, in this case the parametric coefficient G$_{12}$ or G$^*_{12}$, with these two. 

\cite{Oszki2012Icar} used results from her previous work \citep{oszki2011pcs} on the phase curves of about half a million asteroids to study the relationship between the G$_{12}$ parameter and taxonomy. (Note that this work is previous to \cite{penti2016HG} and therefore they did not use the G$_{12}^{*}$.) The authors stated that it could be possible to compute the probability of an object having a given taxon according to its value of G$_{12}$, but that the distributions could be wide and overlapping (see Fig. 1 on their paper).

\begin{figure}[ht!]
\centering
\includegraphics[width=6.0cm]{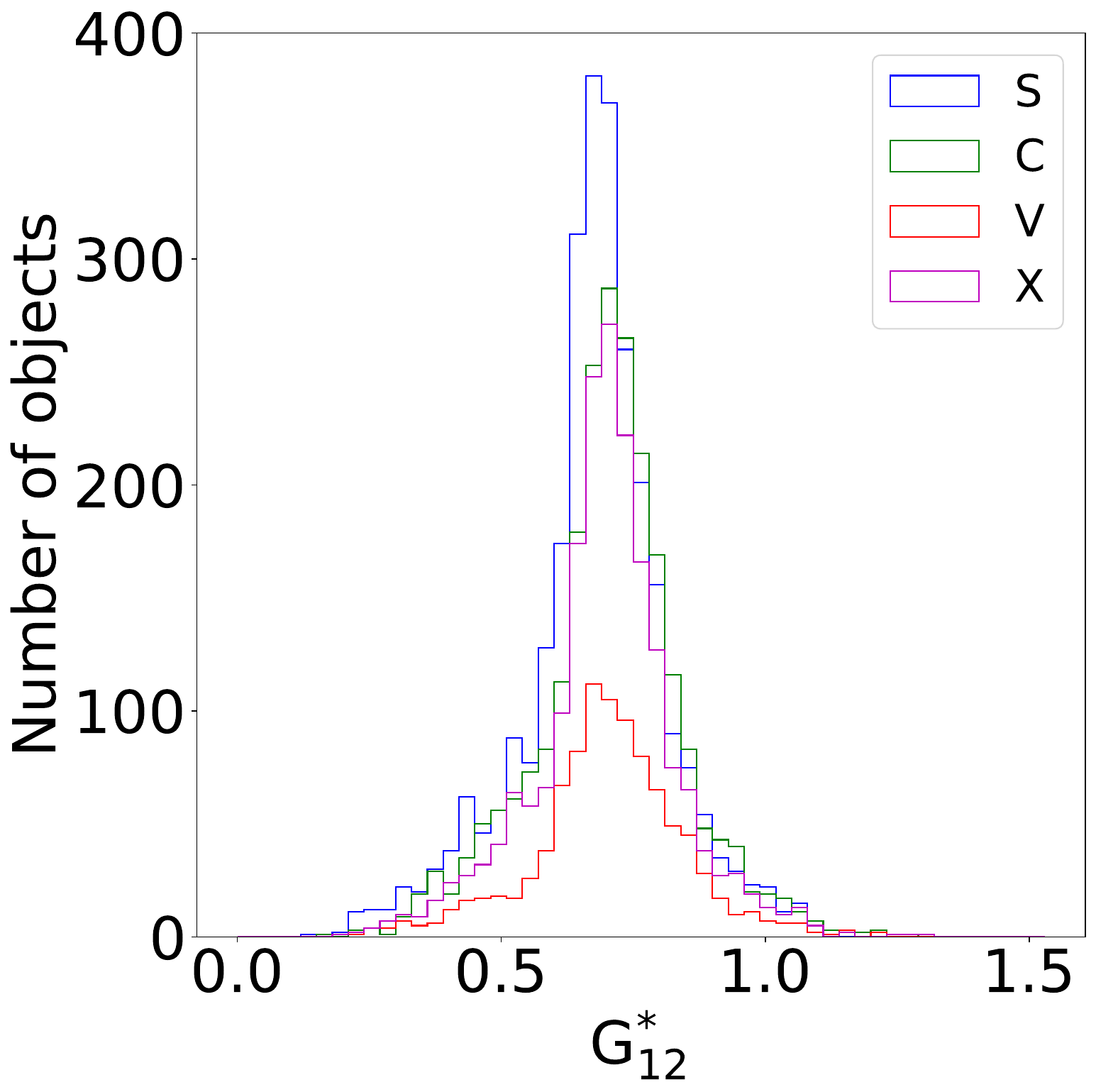}
\includegraphics[width=6.0cm]{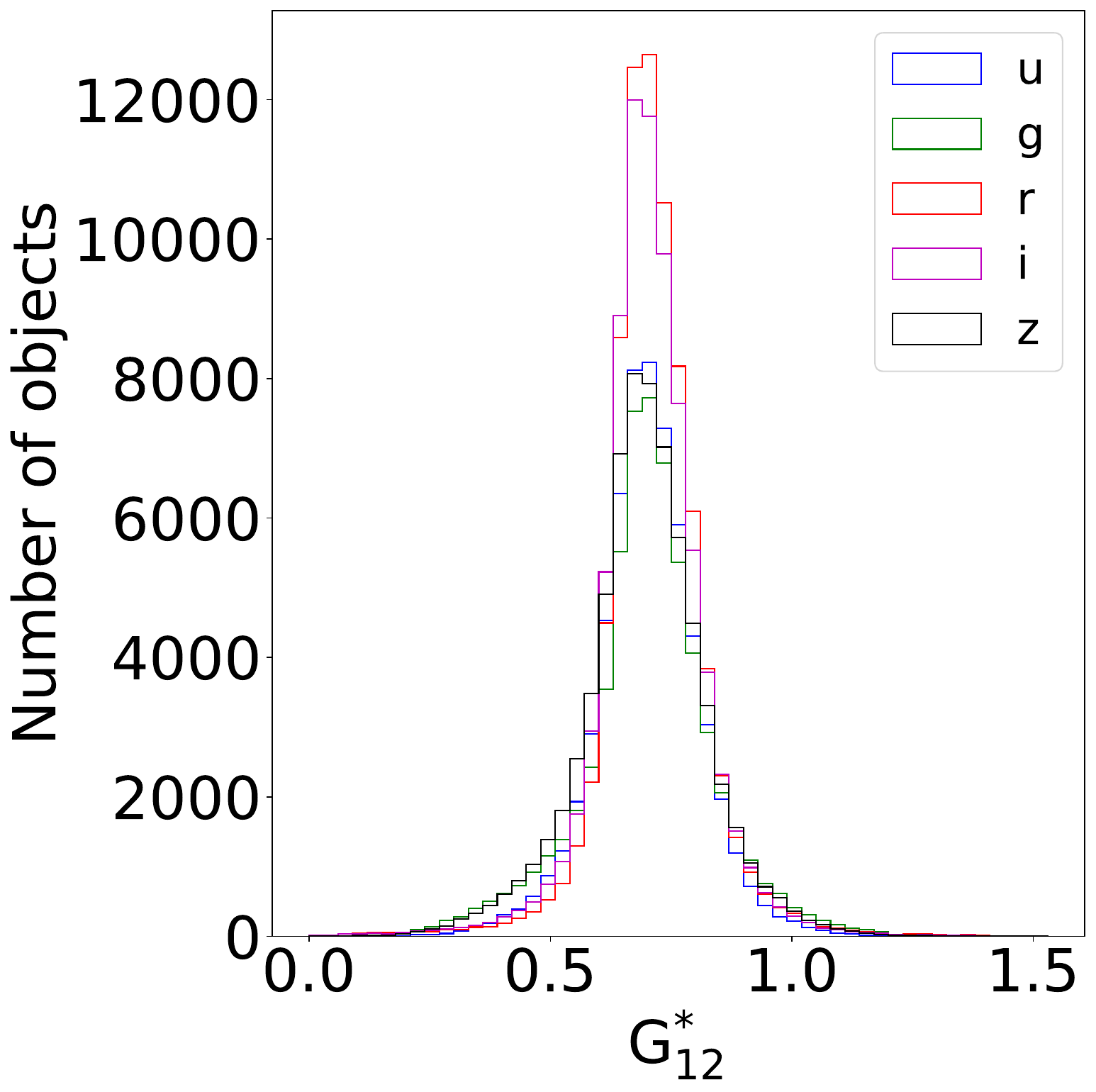}
\caption{Distributions of G$_{12}^*$. Left: The distributions for different taxa are shown in different colors. Right: The distributions for different filters are shown in different colors.}\label{fig:g12}%
\end{figure}

We repeated \cite{Oszki2012Icar} analysis, but using our estimations of G$_{12}^{*}$ \citep[from][]{alcan2024} and the taxonomy classification made by \cite{colazo2022}. {The cross-match between these databases shows in total 3\,196 objects classified as S-complex, 2\,649 as C-complex, 2\,209 as X-complex, and 1\,142 as V-type.} Figure \ref{fig:g12}, left panel, shows that the distributions of G$_{12}^{*}$ are all similar, with mean values of 0.68, 0.70, 0.70, and 0.69 for the S, C, V, and X complexes, as defined in \cite{colazo2022}. The only notable feature in the distribution is the apparent extended tail of the V-complex, but we will not enter into details because this is out of the scope of this work.
We also checked the possible variation of G$_{12}^{*}$ with wavelength but found no evidence of it (see Fig. \ref{fig:g12}, right panel). {The number of objects per filter is shown in Table \ref{tab:numbers}.}

\begin{table}
\centering
\caption{Number of objects. {Columns 1, 4, and 7 indicate the photometric model used; columns 2, 5, and 8 indicate the photometric filter used. Colums 3, 6, and 9 indicate the number of phase curves fitted.}}\label{tab:numbers}
 \begin{tabular}{c | c | c || c | c | c || c | c | c}
 %\hline
 %Model & Filter & Quantity \\
 %\hline
 \hline
HG$_{12}^{*}$ & u   & 61\,711 & Lineal              & u   & 14\,877 & HG$_1$G$_2$         & V   &  3\,819\\
              & g   & 61\,753 & ($\alpha>7.5$ deg)  & g   & 16\,008 &                     &     &        \\
              & r   & 80\,177 &                     & r   & 16\,338 &                     &     &        \\
              & i   & 80\,196 &                     & i   & 16\,331 &                     &     &        \\
              & z   & 68\,622 &                     & z   & 15\,952 &                     &     &        \\
              & all & 42\,168 &                     & all & 17\,387 &                     &     &        \\

 \hline
 \end{tabular}
\end{table}

\subsection{Using G$_1$ and G$_2$}
Using the parametric form of the phase coefficients permits the computation for scores of objects that cannot be fitted using the full {HG$_1$G$_2$  model. The model needs a well-sampled phase curve and data with low-$\alpha$, below 3 degrees, to account for the opposition effect correctly.} Nevertheless, studying what happens when using the model, even if it means working with fewer objects, is helpful.

Since the HG$_1$G$_2$ system was established in 2010 it was proposed a linear relation between the phase coefficients, G$_1$ and G$_2$ \citep[see, for example,][]{muinonen2010HG1G2,shev2016,penti2016HG} where the different taxa more or less fall in different regions of the phase space (C-complex at high G$_1$ and low G$_2$, while S-complex at low G$_1$ and high G$_2$. E-type asteroids break this relation.) {Although recent work by \cite{arcoverde2023MNRAS} on the phase curves of near-Earth asteroids does not show this distribution, showing, instead, a mix of objects with different taxa in different regions, new results obtained using ATLAS sparse data with dense data from ground-based telescopes
support this view \citep{wilawer2024MNRAS}.}

On the other hand, \cite{mahlke2021} computed the phase curves in the cyan and orange filters of the Asteroid Terrestrial-impact Last Alert System, ATLAS, survey \citep{tonry2018PASP-ATLAS} producing several tens of thousands of absolute magnitudes and phase coefficients. They created density plots for different taxa using G$_1$ and G$_2$ (see their work for details). From  Fig. 6 in their paper, it is apparent that, although there are some differences at first glance, all taxa overlap significantly. Moreover, \cite{dobson2023PSJ} fitted phase functions for a dozen transneptunian objects, centaurs, and Jupiter family comets. The authors could use the HG$_1$G$_2$ model for the ATLAS cyan and orange magnitudes for some of these objects. Whenever the values of G$_1$ and G$_2$ are plotted, with uncertainties, on top of \citeauthor{mahlke2021}'s regions, the values are compatible with all different taxonomical classes within one or two $\sigma$, meaning that the statistical difference between them is doubtful.

\begin{figure}[ht!]
\centering
\includegraphics[width=6cm]{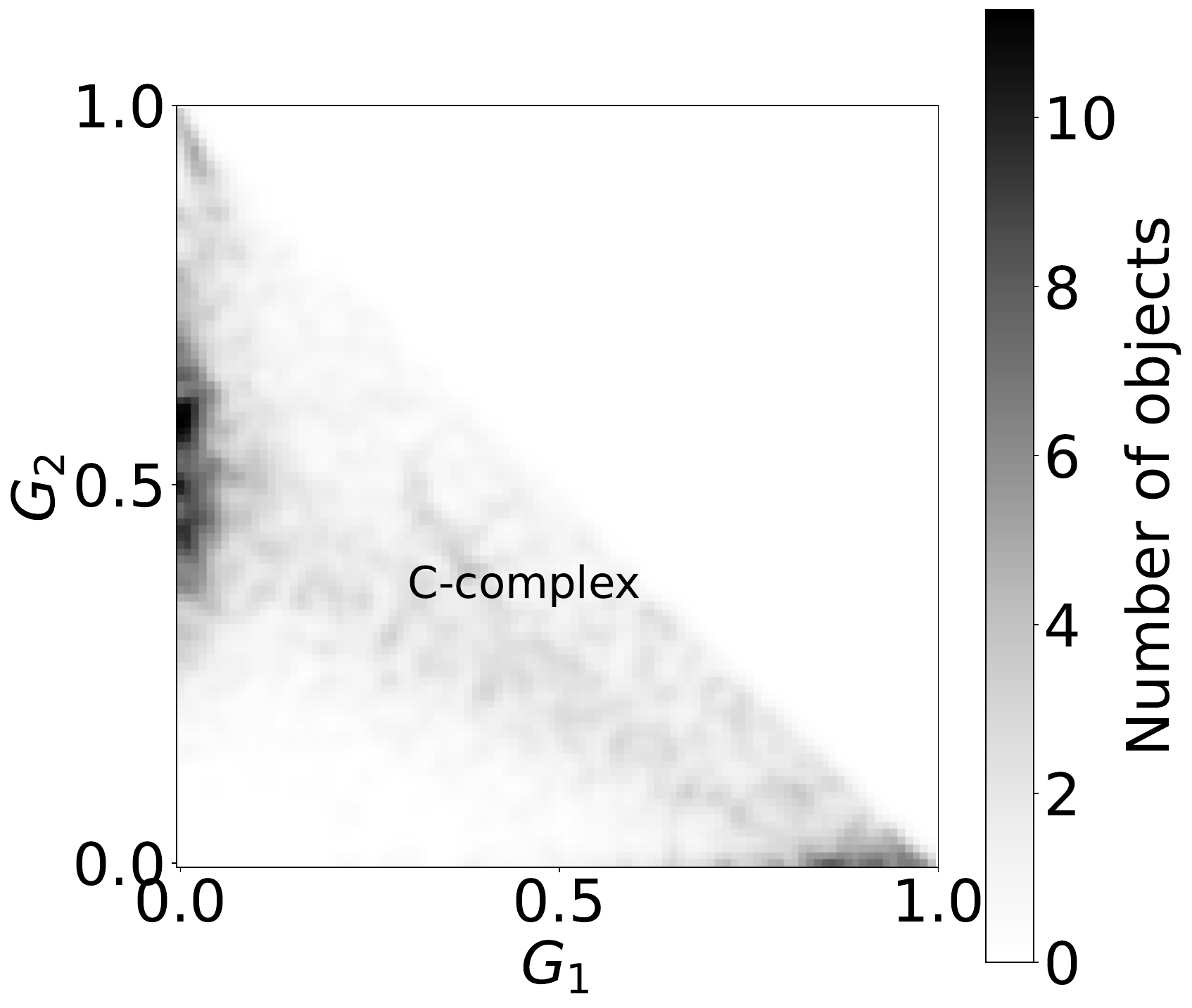}
\includegraphics[width=6cm]{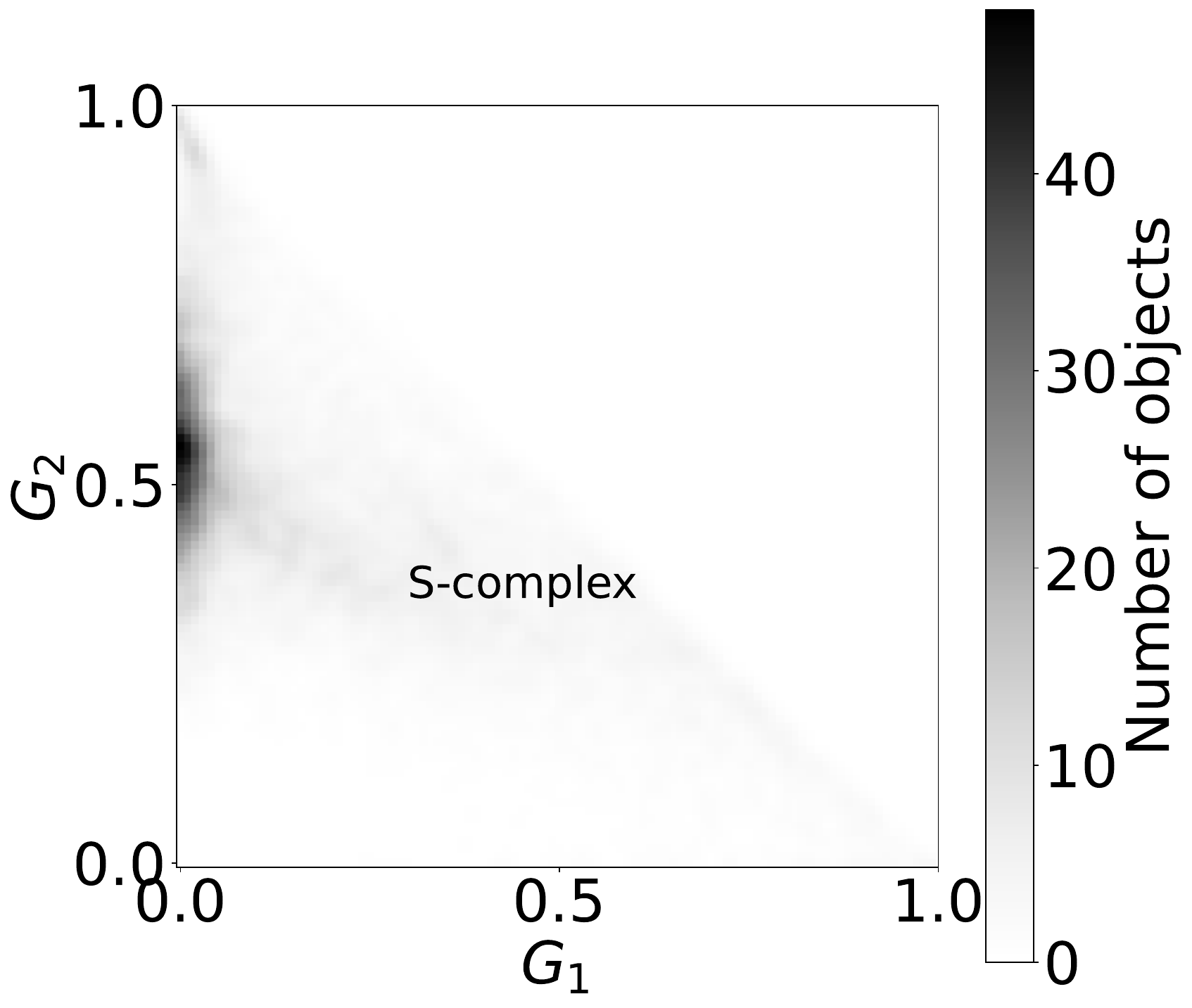}
\includegraphics[width=6cm]{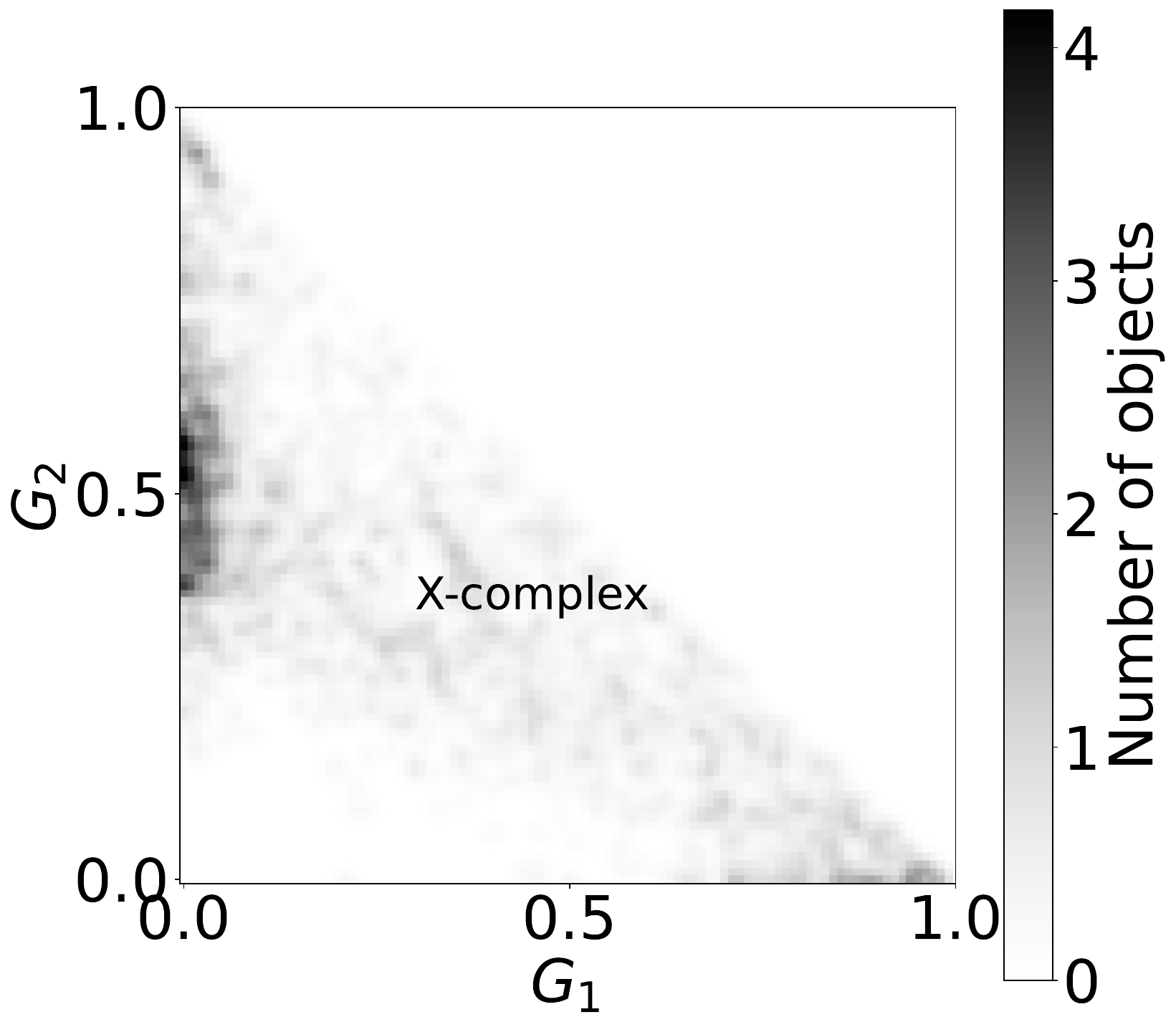}
\includegraphics[width=6cm]{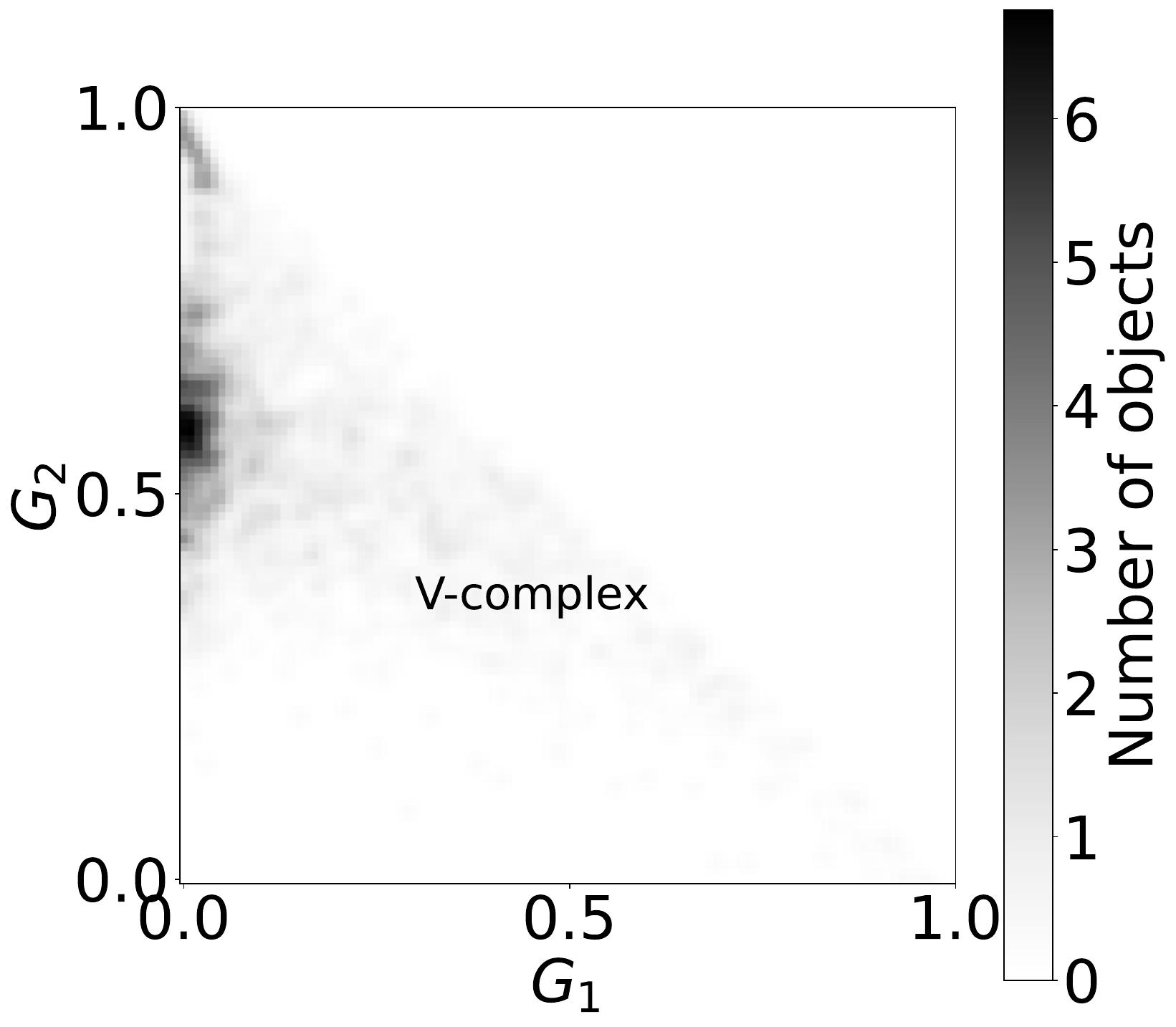}
\caption{Heat maps of G$_1$ versus G$_2$. Each panel shows a different complex (see Table \ref{tab:taxa} for details).}
\label{fig:M21complex}
\end{figure}

To check the accuracy of using G$_1$ and G$_2$ to estimate taxa, we use Mahlke's database, selecting only the values of G$_1$ and G$_2$ obtained with the orange filter (the most numerous) to reconstruct the 2D maps. Instead of using all the taxonomical classes listed in \cite{mahlke2021}, we encompassed them into only four groups: {15\,230 objects in the S-complex, 7\,865 objects in the C-complex, 2\,096 objects in the X-complex, and 2\,009 objects in the V-complex} (see Table \ref{tab:taxa} for details on how Mahlke's taxa were divided into the complexes by \citealt{colazo2022}). {The reader should bear in mind that, due to the overall uncertainties in our data, we needed to make this simplification as we cannot yet detect the level of detail necessary to split our large complexes into smaller groups.}
We used the results from \cite{shev2016} to select a C-type and a S-type asteroid, with values G$_1=0.76\pm0.05$ and $0.30\pm0.03$; and G$_2=0.03\pm0.03$ and $0.30\pm0.03$, respectively. We also included a fictional object with G$_1=0.50\pm0.03$ and G$_2=0.50\pm0.03$ for comparison. Note that these values correspond to phase curves obtained in Johnson's V filter, but for the sake of argument, we will assume they are all compatible with ATLAS' orange filter. Figure \ref{fig:M21complex} shows the phase space covered by the different complexes.

Then, we wanted to answer the question: given the values of G$_i$ of our test subjects, what are the probabilities of falling into one or more complex(es)? Therefore, we normalized each map to the total number of objects per map, and we created two-dimensional Gaussian profiles of each test object using the same binning as shown in Figs. \ref{fig:M21complex} with the values of G$_i$ mentioned above. To compute the probability of a given measurement belonging to each of the complexes, we just multiplied the Gaussian profile (also normalized to the unity area under the 2D curve) to each of the four maps. The results are: For the {\it a priori} C-type asteroid: C-complex 47 \%, X-complex 45 \%, S-complex 6 \%, and V-complex 2 \%; for the S-type asteroid: C-complex 55 \%, S-complex 42 \%, V-type 2 \%, and X-complex 1 \%. The generic asteroid has probabilities V-complex 32 \%, X-complex 26 \%, S-complex 22 \%, and C-complex 20 \%. From these figures, it is unclear that just one pair (G$_1$, G$_2$) could certify a taxonomical classification. However, the high-G$_1$, low-G$_2$ region seems to have a low probability of spectra with the 900 nm absorption band.
\begin{table}
\centering
\caption{Associated taxa}\label{tab:taxa}
 \begin{tabular}{c | c}
 \hline
 Complex${^a}$ & Taxa${^b}$ \\
 \hline
 \hline
 C & Cgh,Ch,B,F,FC,C,CB,CD,CF,CG,CL,CO,Cb,Cg,Cgx,Co,D,DP \\
 S & K,L,LQ,Ld,S,SQ,SV,Sa,Sk,Sl,Sp,Sq,Sqw,Sr,Srw,Sv,Sw,A,AQ,Q,QO,QV\\
 X &M,P,PC,PD,X,XC,XD,XL,Xc,Xe,Xk,Xt,E\\ 
 V & O,V,Vw\\
 \hline
 \end{tabular}
 {\small ${^a}$ {Adapted from \cite{colazo2022}}\\ 
 ${^b}$ {From \cite{mahlke2021}}}
\end{table}

\subsection{Geometrical albedoes and phase coefficients}

Another parameter that enters into the taxonomic classification is the geometrical albedo of the objects. It is well-established that the C-complex and some X-complex objects tend to have low albedoes, while the S and V-complexes are associated with higher albedoes. Therefore, we checked whether using the albedo as an extra parameter in the G$_1$-G$_2$ space could distinguish among different kinds of objects.

\begin{figure}[ht!]
\centering
\includegraphics[width=6cm]{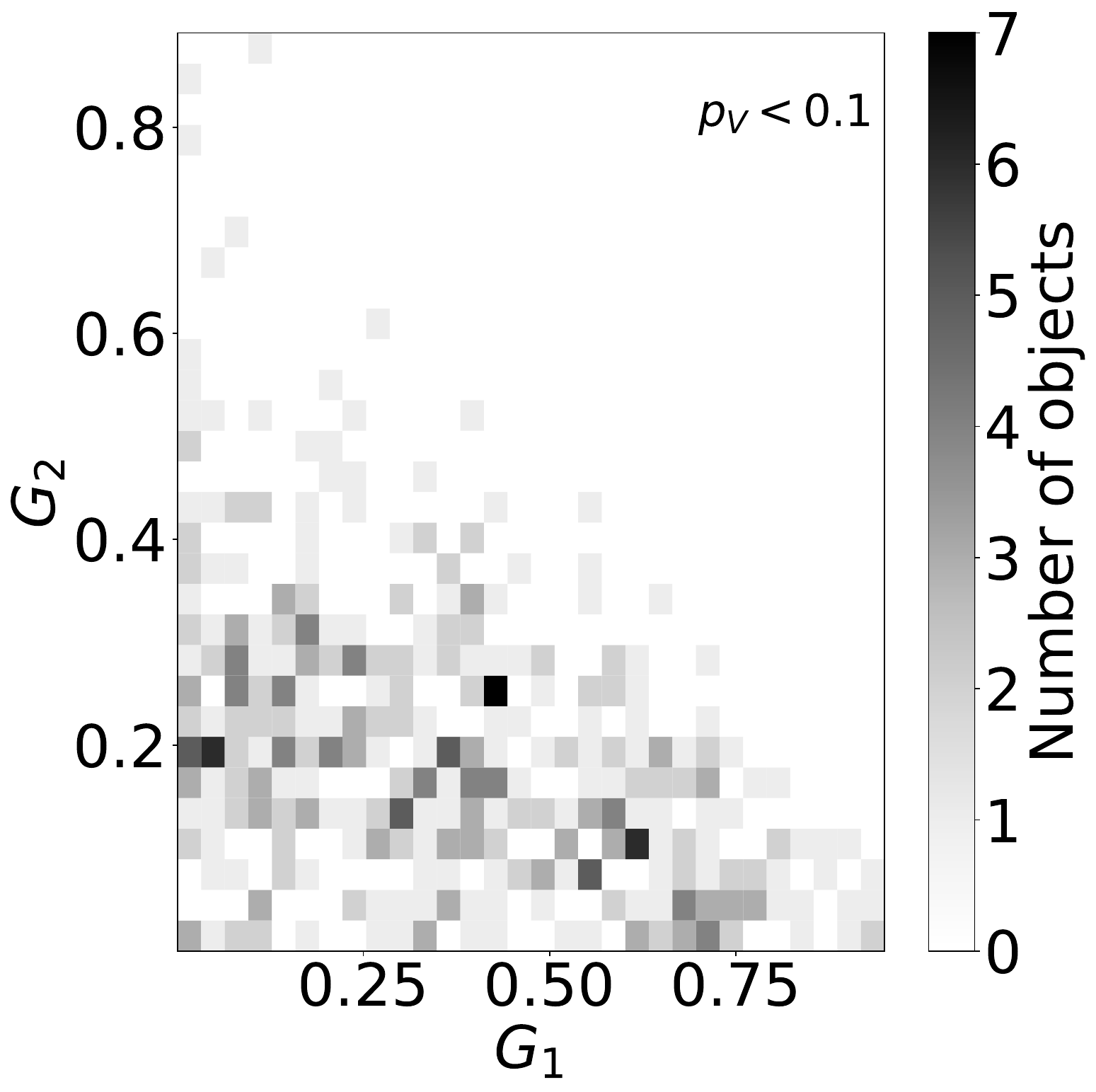}
\includegraphics[width=6cm]{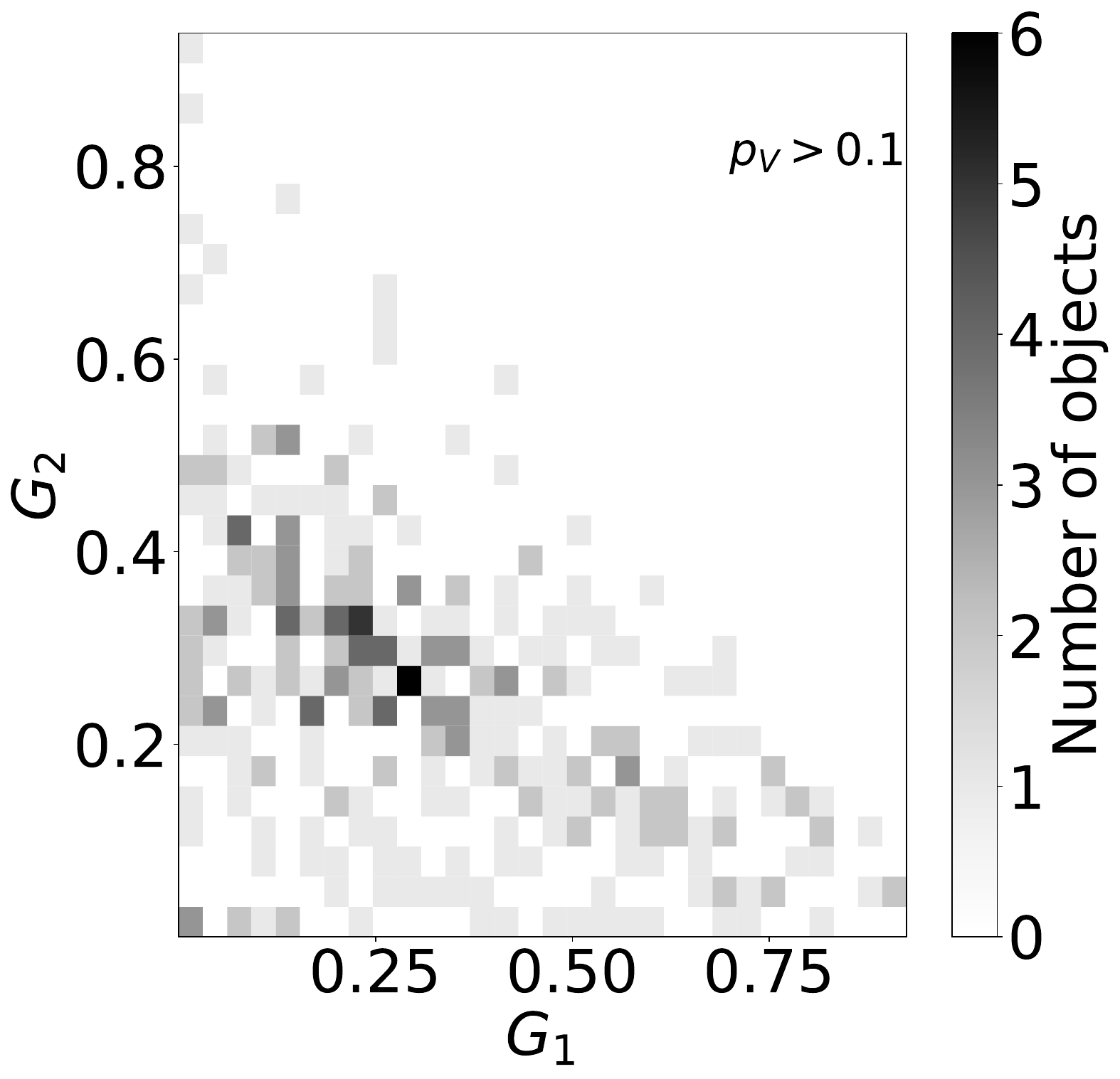}
\caption{Heat map of G$_1$ versus G$_2$. The left panel shows objects with $p_v<0.1$, while the right panel shows $p_v>0.1$. Albedoes extracted from the WISE database.}
\label{fig:albedosg1g2}
\end{figure}

We used our values of G$_1$ and G$_2$ computed from the data in \cite{sergeyev2021AyA}. We applied the HG$_{1}$G$_2$ model to all objects with at least three data, spanning a minimum of five degrees, and used the same methodology as in \cite{alcan2024} {only for the V filter}. We obtained results for almost 4\,000 objects\footnote{In this case, we did not compute the complete probabilities distributions, but we kept the nominal values to save computing time. The data is in the Open Science Framework, OSF, folder https://osf.io/v2wkj/.} ({see Table \ref{tab:numbers}}). We cross-matched our results with the WISE albedoes \citep{wisealbedo2019PDSS}. The cross-match provided almost 800 matches. The results are display in Fig. \ref{fig:albedosg1g2} separated in two samples: low-albedo ($p_v<0.1$) and high-albedo ($p_v>0.1$).
There is no clear distinction among different albedoes in the figures, except perhaps a concentration of higher-albedoes objects towards values of G$_1<0.5$, but the distribution generally seems reasonably uniform. {In any case, these results should regarded with care because most of the phase curves have few points; nevertheless, there are about 350 objects with more than ten observations. Furthermore, on average, the results should not differ much from higher coverage phase curves, but a more thorough analysis is warranted when higher-quality data is available.}

\subsection{Relationship between albedo and the linear part of the phase curve}

Regarding the use of albedoes and phase curve parameters, a significant relation that has been explored since \cite{belskayashev2000} put it forward is the apparent correlation found between the slope of the linear part of the phase curve, $h$, and albedo. The phase curves of small bodies have a linear behavior for $\alpha\gtrsim7$ deg. The (anti) correlation seems like a strong result in the literature (see Fig. 4 in their work or, also Fig. 4, in \citealt{wilawer2022mnras}). Nevertheless, neither of these references reports the result of an actual correlation test. Moreover, scrutiny to \cite{wilawer2022mnras} data in Table 2 shows objects departing from apparent anticorrelation. These objects, the let-outs, have uncertainties more significant than 20 \%, which are less precise values but not necessarily less accurate ones.

\begin{figure}[ht!]
\centering
\includegraphics[width=6cm]{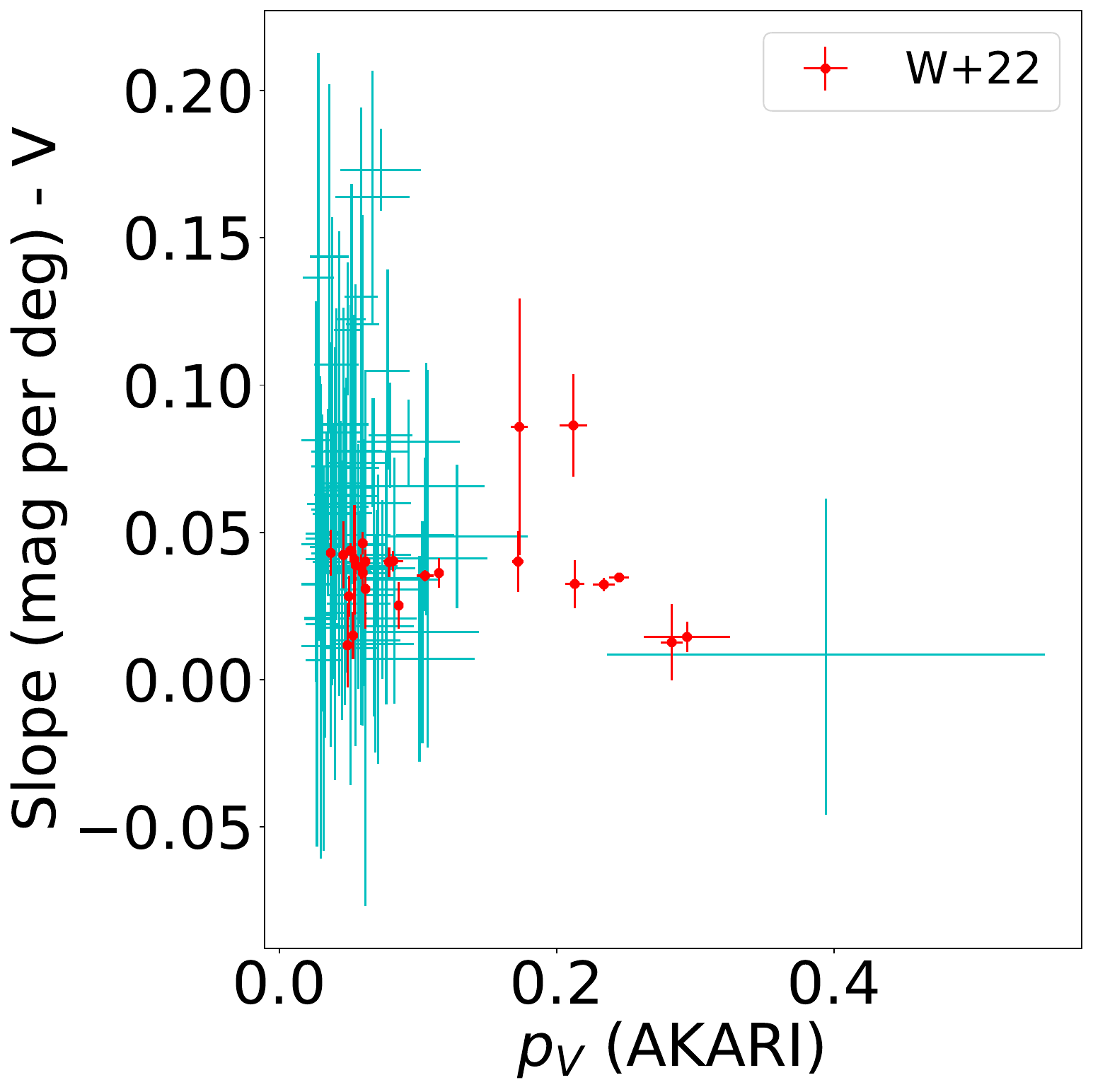}
\includegraphics[width=7.1cm]{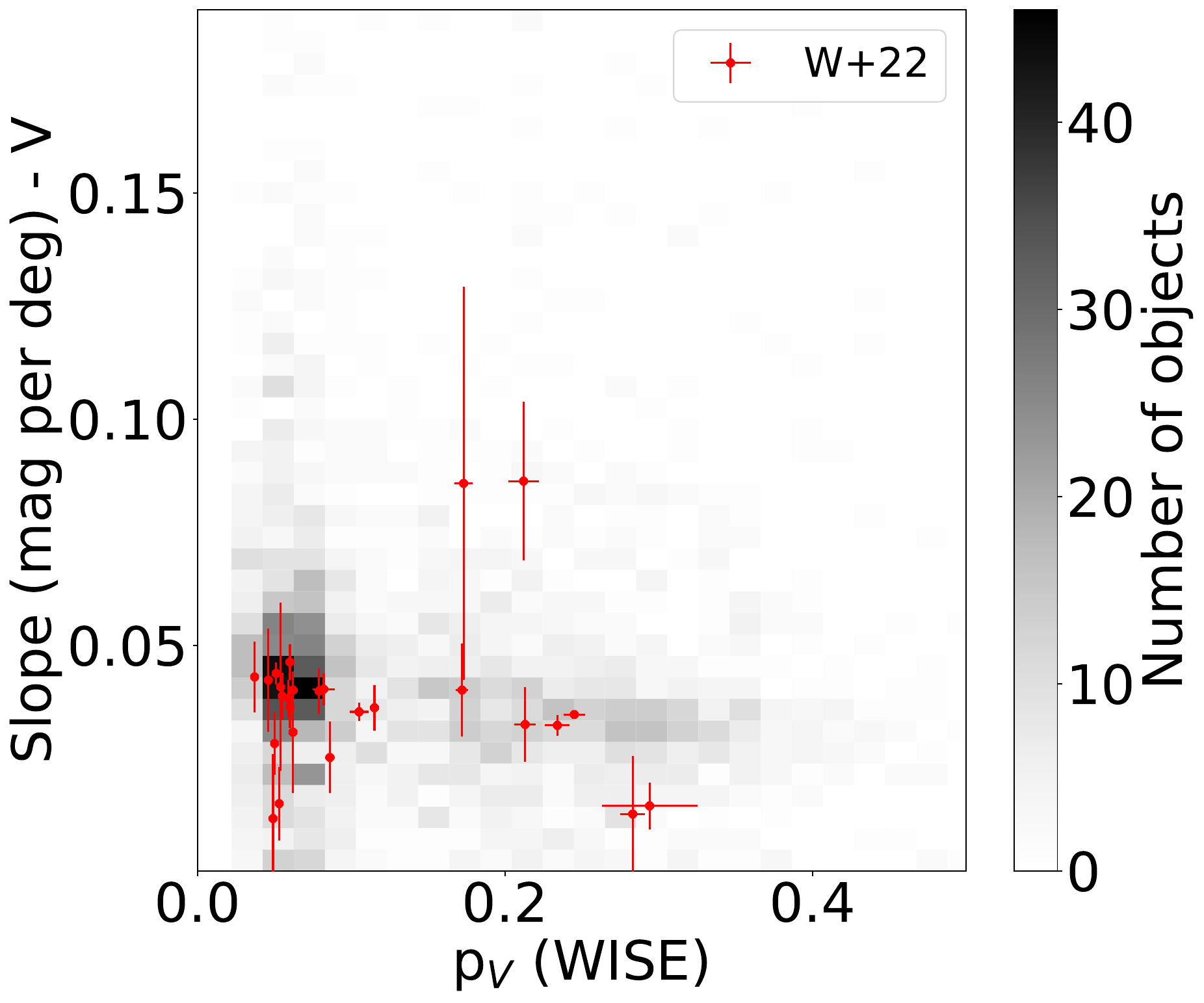}
\caption{Left: Geometric albedo versus $h$. Albedoes from the AKARI mission. Right: Heat map of geometric albedo versus $h$. Albedoes from the WISE mission. The plot was cut for $p_v<0.5$ for clarity. Overplotted is \cite{wilawer2022mnras}'s data in red.}
\label{fig:slope_alb}
\end{figure}

In Fig. \ref{fig:slope_alb}, we show all objects in \cite{wilawer2022mnras} as they appear in their Table 2; in these cases of objects with more than one value, we selected the lower uncertainty to be as close as possible to their original criterion. The correlation disappears. Moreover, Spearman's test shows that the null hypothesis cannot be rejected, i.e., that the quantities are uncorrelated.

A fair question is, what would happen if more data is added to the plot? Will the relation hold, dilute, or disappear? As in the previous Section, we used \cite{sergeyev2021AyA}'s data to compute the $h$ for as many objects as possible. The selection criteria we used involved phase curves with at least 3 data, $\alpha_{min}>5.0$ deg, and a minimum span of 5 deg. We only used $m : \sigma_m\leq1$ and the same method as in \cite{alcan2024} but using a linear function as a photometric model. {See the total numbers in Table \ref{tab:numbers}.} The outputs are the probability distributions of the slopes ($h$). The distributions of the almost 24\,000 objects are available in the OSF folder {https://osf.io/v2wkj/}.

Using our $h$, we first matched the AKARI albedo list \cite{alilagoa2018} to our list, finding only 83 objects in common (see Fig. \ref{fig:slope_alb}, left panel). We over-plot the data from \cite{wilawer2022mnras} as red points in the figure. Note that in the figure are only plotted objects with $h\leq0.2$ mag per degree to match the range span from \cite{wilawer2022mnras}. The Spearman test does not show evidence of correlation.
Still, the AKARI albedos do not cover the phase space Wilawer's data covers. Note that only one datum has an albedo larger than 0.2.

\begin{figure}[ht!]
\centering
\includegraphics[width=7.1cm]{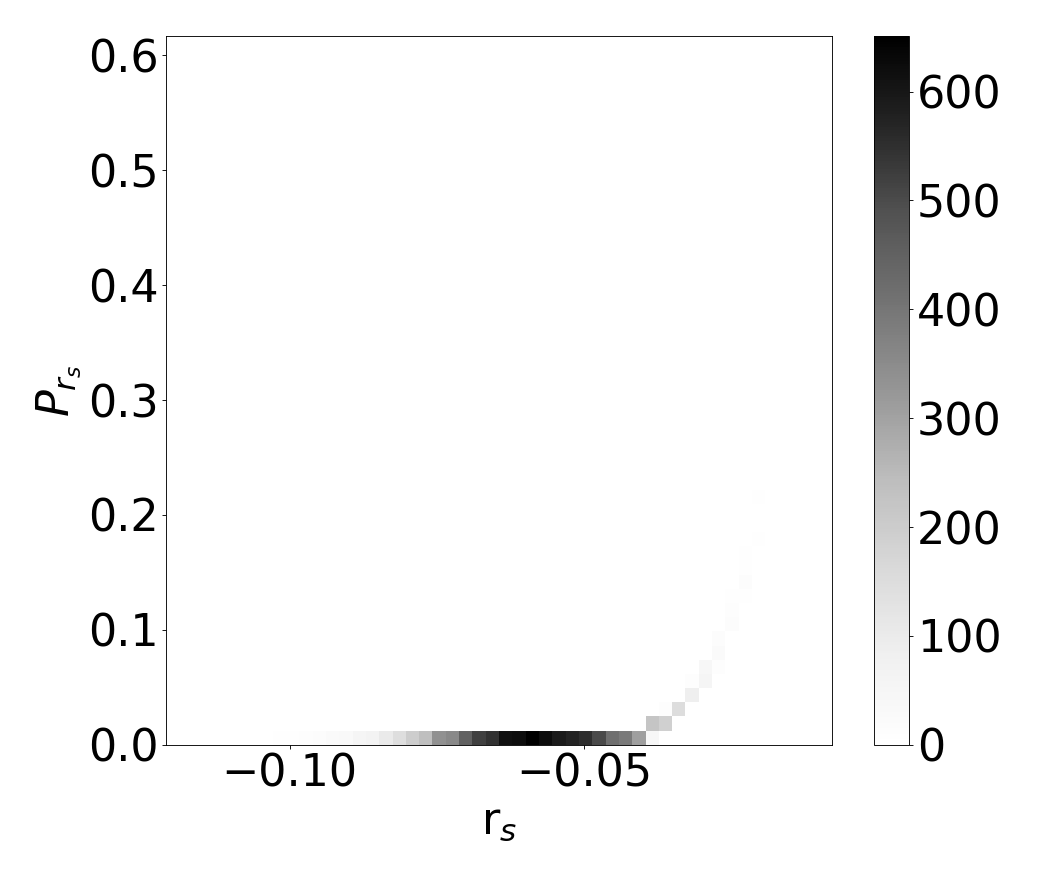}
\caption{Distribution of the 10\,000 Spearman test' runs.}
\label{fig:corr_test}
\end{figure}

Next, aiming at increasing the numbers, we cross-matched our $h$ with the albedos from the WISE mission \citep{wisealbedo2019PDSS}, obtaining 4\,122 matches. The comparison is shown in Fig. \ref{fig:slope_alb}, right panel, where the data is displayed as a heat map to ease visualization. {In this case, the Spearman test shows a weak anticorrelation with a $r_s = -0.2$. To explore how the uncertainty of the data may affect this anticorrelation, we generated 10\,000 clones using normal distributions with $\sigma$ equal to the uncertainties in the $x$ and $y$ axes. The results of $r_s$ and $P_{r_r}$ are shown in Fig. \ref{fig:corr_test} and indicate that a weak anticorrelation, most of the results cluster around $p_r\approx-0.06$ with very low $P_{r_r}$, may exist, implying steeper face curves tend to have lower-albedoes.}

\section{Taxonomy based on absolute colors}

The above issues were raised not in an attempt to minimize the importance of previous (nor future) knowledge but to show that the upcoming flood of data, such as that from the Rubin Observatory's Legacy Survey of Space and Time, LSST, will open up new avenues for improvement in the way we interpret data. This influx of data will not only change the way we look at current results but also provide the potential to reinterpret long-standing relations that may become blurred when more (a lot more) data is added.

In \cite{alcan2022}, we presented absolute colors for over 14\,000 objects. One of the objectives was to measure multi-wavelength absolute magnitudes to obtain the taxonomic classification of the objects \citep[$>8\,000$ objects in][]{colazo2022}. We extended this work for $>40\,000$ objects in \cite{alcan2024}, but a taxonomical classification is yet to be done. Indeed, this number is about only 10\% of what we would get if using average colors or some other sorting algorithm. Therefore, is it worth the trouble to compute phase curves in multiple wavelengths? The answer is yes because the absolute colors are corrected by phase coloring, thus removing one possible bias. {However, there may be issues when collating data from different apparitions. Nevertheless, the method used in \cite{alcan2024} somehow includes this while computing the probability distribution $P_A(m,\Delta m)$ (see their Eqs. 3 and 4). It can be further improved\footnote{Alternative methods, such as the modified sHG$_1$G$_2$ \citep{Carry2024} deal with these issues with a different approach.}}. Moreover, we have shown in \cite{alcan2022} and \cite{alcan2024} that using average colors biases the results towards low-$\alpha$ observations and that there is not a 1-1 relation between absolute colors and average colors; in fact, there is quite a large spread. Finally, in \cite{alcan2024}, we showed that the spectral slopes and colors of the objects do not follow the traditionally assumed reddening tendency with increasing $\alpha$. Moreover, two different kinds of photometric behavior that change at a critical angle of about 5 deg seem to exist {(see also \citealt{wilawer2024MNRAS})}. These behaviors are unrelated in any straightforward way to the actual taxa of the objects, size, or location in the Solar System.

As a side product, the absolute colors (or the respective relative reflectances) can be used as templates to study the phase effects because they provide the zero-phase angle benchmark and the effects of space weathering, providing uncolored templates.

\section{Conclusions}

This article highlights that estimating taxa from phase coefficients or relations based on {small numbers of objects} may not be ideal. {The main problem when using phase curves is the need for more than one, in fact, a lot more than one, observations of the same object at different $\alpha$. Therefore, using phase angle-corrected colors to estimate taxa is less efficient than using single observations or averages, which do not have the risk of failing to fit a photometric model. Nevertheless, the near future seems bright, as the arrival of the LSST and its synergy with other large photometric surveys with similar photometric systems (for example the Dark Energy Survey, DES, \citealt{DES2016MNRAS}, the Javalambre Physics of the Accelerating Universe Survey, J-PAS, \citealt{benitez2014arXiv,bonoli2021mjpas}, or the SkyMapper's Southern Sky Survey, \citealt{wolf2018SM}) will provide for observations of scores of objects in different $\alpha$ and epochs. Nevertheless, it is important to remember that, although similar, the photometric systems are not identical. Therefore, conversion relations must be used or developed, increasing the uncertainty budget.}

Using absolute magnitudes to estimate taxa should eliminate some of the blur when different objects' observations are used. There are some limitations to the method we have been developing; for instance, it tends to have larger uncertainties than traditional ones, such as averaging. However, we deal with complete probability distributions containing, if not all, at least most of the possible solutions within the current uncertainties of the observational data and rotational state of the object. As mentioned before, including more data from LSST or any other survey with a similar photometric system, or whose data can be transformed into SDSS-like, should improve the posterior distributions, increasing the precision of our absolute magnitudes, thus improving the taxonomical classification, which is free of phase angle effects.

\section*{Acknowledgments}
\noindent
I thank K. Muinonen's comments and an anonymous reviewer, who helped me think a lot about the results presented here.
AAC acknowledges financial support from the Severo Ochoa grant CEX2021-001131-S funded by MCIN/AEI/10.13039/501100011033. 
All figures in this work can be reproduced using the Google Colab Notebook https://colab.research.google.com/drive/1Qob-Zz7ZFpf0g0Oe0yjfCQl9l1qGgG\_q?usp=sharing.\\

%% The Appendices part is started with the command \appendix;
%% appendix sections are then done as normal sections
%\appendix
%section{Example Appendix Section}
%\label{app1}

%Appendix text.

%% For citations use: 
%%       \citet{<label>} ==> Lamport (1994)
%%       \citep{<label>} ==> (Lamport, 1994)
%%
%Example citation, See \citet{lamport94}.

%% If you have bib database file and want bibtex to generate the
%% bibitems, please use
%%
\noindent
{\it Declaration of generative AI and AI-assisted technologies in the writing process:
During the preparation of this work the author used Grammarly in order to revise the English of this manuscript. After using Grammarly, the author reviewed and edited the content as needed and take full responsibility for the content of the publication.}

\bibliographystyle{elsarticle-harv} 
\bibliography{elsarticle-template-harv}

%% else use the following coding to input the bibitems directly in the
%% TeX file.

%% Refer following link for more details about bibliography and citations.
%% https://en.wikibooks.org/wiki/LaTeX/Bibliography_Management

%\begin{thebibliography}{00}

%% For authoryear reference style
%% \bibitem[Author(year)]{label}
%% Text of bibliographic item

%\bibitem[Lamport(1994)]{lamport94}
%  Leslie Lamport,
%  \textit{\LaTeX: a document preparation system},
%  Addison Wesley, Massachusetts,
% 2nd edition,
%  1994.

%\end{thebibliography}
\end{document}